\documentclass[12pt]{article}
\usepackage{hyperref}
\usepackage{graphicx}
\usepackage{amssymb,amsmath}
\usepackage{epstopdf}
\usepackage{color}
\renewcommand\[{\left[}

\def\be{\begin{equation}}
\def\ee{\end{equation}}
\def\bea{\begin{eqnarray}}
\def\eea{\end{eqnarray}}

\def\bea{\begin{eqnarray}}
\def\eea{\end{eqnarray}}

\title{
\begin{flushright}
\small
UAB-FT-598\\FTUAM-06-02\\IFT-UAM/CSIC-06-10\\DESY-08-167\\MPP-2008-158\\GEF-TH-4-08
\end{flushright}
\vspace{.5cm}
{\Large\textbf{Mixing of photons with massive spin-two particles\\ in a magnetic field}} }

\author{
Carla Biggio$^{1,2,3}$, 
Eduard Mass{\'o}$^{4}$ and 
Javier Redondo$^{4,5}$
\\[2ex]
\small{ $^1$Departamento de F\'\i sica Te{\'o}rica, Universidad Aut{\'o}noma de Madrid, Spain}\\
\small{ $^2$Max-Planck-Institut f\"ur Physik, M\"unchen, Germany}\\
\small{ $^3$Dipartimento di Fisica, Universit\`a di Genova, Italy}\\
\small{ $^4$Institut de F\'\i sica d'Altes Energies, Universitat Aut{\`o}noma de Barcelona, Spain}\\
\small{ $^5$Deutsches Elektronen-Synchrotron DESY, Hamburg, Germany}
}
\date{}

\begin{document}

\maketitle

\begin{abstract}
We study the mixing of photons with hypothetical massive spin-two
particles in the presence of a magnetic field. 
Mixing phenomena have been studied in the case of axion-like particles 
and strictly massless spin-two particles (gravitons) but not in this case. 
We find several interesting differences between them.
\end{abstract}


\section{Introduction}
\label{introduction}

A photon traveling through an electromagnetic field may convert into
neutral particles that have a coupling to two photons. The process has
been observed in the laboratory; indeed, photons in the electric field
of a nucleus produce pions as predicted by
Primakoff~\cite{Primakoff:1951ww}.  If the particle is light enough we
expect that the mixing of the photon with the particle leads to a
coherent superposition of the two.  Such an effect is at the heart of
the proposals to look for axions, since coherence enhances the
probability of axion-photon conversion~\cite{Sikivie:1983ip}. All
current axion searches are based on detection techniques that exploit
the increase in the sensitivity that this mixing phenomenon
provides. These searches include haloscopes~\cite{Asztalos:2001tf},
helioscopes~\cite{Lazarus:1992ry,Moriyama:1998kd,Inoue:2002qy,Inoue:2008zp,Zioutas:2004hi,Andriamonje:2007ew} 
and laser experiments~\cite{Ruoso:1992nx,Cameron:1993mr,Zavattini:2005tm,Zavattini:2007ee,Chou:2007zzc,Fouche:2008jk,Chen:2006cd,Ehret:2007cm,Afanasev:2008jt,Pugnat:2007nu,Cantatore:2008ju}
(for a review see Ref.~\cite{Eidelman:2004wy}).

Since the axion is a theoretically well motivated particle, almost all
mixing studies have focused on this case.  Of course any other light
spinless particle coupling to photons could lead to the same
phenomenology~\cite{Masso:1995tw}.  Furthermore, photons may oscillate
also into particles with spin higher than one (mixing with spin-one
particles is forbidden due to the Landau-Yang
theorem~\cite{Landau:1948,Yang:1950rg}), the most prominent case being
the graviton.

Photon-graviton mixing has being considered in the framework of
standard four-dimensional general
relativity~\cite{Gertsenshtein:1961,Aichelburg:1970dh,Magueijo:1993ni,Bastianelli:2004zp,Bastianelli:2007jv,Raffelt:1987im},
where gravitons are massless, and also in the context of
extra-dimensions~\cite{Deffayet:2000pr} where, in addition to the
massless mode, one finds a Kaluza-Klein tower of massive gravitons.
Unfortunately, due to the extreme smallness of Newton's constant, the
effects of photon-graviton mixing are usually of little relevance.

In this paper we want to present our calculations for photon
mixing with a generic massive spin-two particle, in principle
unrelated to Kaluza-Klein gravitons.

In our study we have seen that there are some peculiarities in the
spin-two case that are not present when the mixing is with
pseudoscalars such as axions; we consider instructive to present such
differences.  Even more than that, we have realized that even if the
mass $m$ of the spin-two particle is small, there are significant
differences with the strictly massless case.  Specifically, we find
that there is no decoupling, in the sense that in the $m \rightarrow
0$ limit there are effects not present in the massless theory. The
helicity-zero component of the massive theory contributes without
suppression to the observables we calculate.

Our results may be useful for a certain number of
experiments on laser propagation in magnetic fields~\cite{Ruoso:1992nx,Cameron:1993mr,Zavattini:2005tm,Zavattini:2007ee,Chou:2007zzc,Fouche:2008jk,Chen:2006cd,Ehret:2007cm,Afanasev:2008jt,Pugnat:2007nu,Cantatore:2008ju}. 
Indeed, photon mixing leads to several effects and it is of interest whether
we may interpret or not such potential effects as the result of the
mixing of photons with spin-two particles.  Again, the finding related
to the contribution of the helicity-zero component makes this
interpretation as natural as the spinless one, in the potential case
that effects are found experimentally.  However, one should always
bare in mind that these couplings are subject to strong constraints
not only from
astrophysics~\cite{Lazarus:1992ry,Moriyama:1998kd,Inoue:2002qy,Inoue:2008zp,Zioutas:2004hi,Andriamonje:2007ew,Raffelt:1996wa,Raffelt:1999tx,Schlattl:1998fz},
which is actually the case of axions, but also from the usually very
demanding laboratory searches for long range
forces~\cite{Adelberger:2006dh,Schlamminger:2007ht,Dupays:2006dp}. Some
refined models that evade constraints from astrophysics and/or
searches for new long range forces have been
developed~\cite{Masso:2005ym,Brax:2007ak,Masso:2006gc,Mohapatra:2006pv},
but they all suffer from some lack of naturalness.

Last but not least our work might be of interest concerning the
controversy of the massless limit of a theory of massive gravity
(see~\cite{Goldhaber:2008xy} for a recent review). The issue goes back
to the study of van Dam and Veltman~\cite{vanDam:1970vg} and
Zakharov~\cite{Zakharov:1970cc}, who showed that there was a
discontinuity in the limit of the graviton mass going to zero when
compared with the usual gravitational theory with a massless graviton,
and to the study of Vainshtein~\cite{Vainshtein:1972sx}, who argued
that the non-linearity of Einstein equations could solve the
problem. There has been much work on such topic and, as far as we
know, there is no clear consensus.  It is not our intention to enter
into the intricacies of such deep problems, but rather to present a
calculation that shares with gravitation the fact that we also deal
with a spin-two particle. In our case, which by the way is developed
in linear theory, we find a discontinuity in the $m \rightarrow 0$
limit.

In this paper we shall not review the pseudoscalar case; the work of
Raffelt and Stodolsky~\cite{Raffelt:1987im} is very complete and can
be complemented with more recent
studies~\cite{Hoogeveen:1990vq,Adler:2008gk}.  Instead we will present
in Section~\ref{scalar} the mixing of photons with scalar
particles. This will help us in fixing some of the notations and
conventions we use. Also, another motivation is that there are small
technical differences between the scalar and the pseudoscalar cases,
unnoticed in previous literature~\cite{Maiani:1986md,Liao:2007nu},
which will also appear in the spin-two calculation.
Section~\ref{tensor} is devoted to the spin-two calculation and a
final discussion is presented in Section~\ref{discussion}. Some
technical details on how we choose the polarization basis and the
demonstration of an useful identity are shown in the appendixes.


\section{Mixing of Photons with Scalars}
\label{scalar}

The Lagrangian describing a scalar field $\phi$ coupled to two photons
with coupling constant $g_0$ is
\begin{equation}
\label{lzero}
{\cal L}={\cal L}_{EM}+{\cal L}_{KG}+\frac{1}{4}g_0\
F^{\mu\nu}F_{\mu\nu}\, \phi \ ,
\end{equation}
where $F_{\mu\nu}$ is the electromagnetic field strength and we have
the usual free parts
\bea
{\cal L}_{EM}&=&-\frac{1}{4}F^{\mu\nu}F_{\mu\nu}\\
{\cal L}_{KG}&=&
\frac{1}{2}(\partial_\alpha \phi)(\partial^\alpha
\phi)-\frac{1}{2}\,m_0^2\,\phi^2 \ .
\eea
We use the metric
$\eta_{\mu\nu}=\rm Diag\{1,-1,-1,-1 \}$. Following the mixing formalism of
Ref.~\cite{Raffelt:1987im}, we decompose $F_{\mu\nu}$ as
\begin{equation}
\label{Fmunu}
F_{\mu\nu}=F_{\mu\nu}^{ext}+\partial_\mu
A_\nu-\partial_\nu A_\mu \ ,
\end{equation}
where $F_{\mu\nu}^{ext}$ represents the external field and $A_\mu$ the
propagating quantum photon field. The mixing equations in the Lorentz
gauge and considering the external magnetic field constant are:
\begin{eqnarray}
\label{eom} (\partial^2+m_0^2)\phi \  &=& \ \frac{1}{2}g_0\
F^{ext}_{\mu\nu}\ (\partial^\mu
A^\nu-\partial^\nu A^\mu) \\
\partial^2 A_\nu  \ &=& \ g_0\ 
F^{ext}_{\mu\nu}  \ \partial^\mu \phi \ .
\label{eom-second}
\end{eqnarray}
We choose the $z$-axis as the direction of propagation of light and
the magnetic field as $\vec
B=(B_T,0,B_L)=(F^{ext}_{23},0,F^{ext}_{12})$.  We also have
$A_\mu=(0;\vec A) =(0;A_{||},A_{\perp},0)$, with polarizations
$A_{||}$ and $A_{\perp}$ respectively parallel and perpendicular to
$B_T$.

We will work in the case in which $m_0$ is small and the magnetic
field is stationary and slowly varying in space, which are the
conditions typically interesting for laboratory experiments. The
motivation of the first requirement is two-fold. On the one hand $m_0$
has to be smaller than the photon energy because otherwise real
$\gamma-\phi$ transitions can not occur. On the other hand if
$m_0^2/2\omega^2\ll 1$ ($\omega$ is the energy) we can linearize the
equations of motion, which simplifies the calculation without loosing
relevant information.  Moreover, we consider $g_0$ small enough so
that we can work at first non-trivial order~\footnote{From an
effective field theory point of view, we can consider the
$\phi\gamma\gamma$ coupling in Eq.~(\ref{lzero}) as being generated by
new physics related to an energy scale $\propto g_0^{-1}$.  Therefore,
going beyond the lowest non-trivial order in $g_0$ would be probably
inconsistent since, whatever the new physics is, it will most likely
generate other effective interactions with couplings proportional to
higher powers of $g_0$ which would have comparable impact on the final
results.}.

With all these conditions satisfied, the field solutions to the
equations of motion in the stationary regime will be plane waves with
an energy dependence $ e^{i \omega t }$ and a spatial dependence
$e^{-i p z}$ to be determined~\footnote{Given the way we chose the
axes, $x$ and $y$ play no role.}.  In the relativistic case in which
$m_0\ll \omega$ we expect $|p|\simeq \omega$ and we can linearize the
operator $\partial^2 $ in the equations of motion:
\begin{equation}
\label{lineal}
\omega^2+\partial_z^2 =
(\omega+i\partial_z)(\omega-i\partial_z)=
(\omega+p)(\omega-i\partial_z) \simeq 2\omega(\omega-i\partial_z) \ .
\end{equation}
By reducing the order of the system of differential equations to one
we are loosing the possibility of imposing boundary conditions on the
field derivatives and thus of considering reflected waves.  In turn
this can be used to argue that the amplitudes of such waves should be
of the order of the neglected terms when we linearize as in
Eq.~(\ref{lineal}), i.e. suppressed by a factor $(\omega-p)/\omega$ at
least. Explicit formulas for the case of the axion can be found
in Ref.~\cite{Hoogeveen:1990vq,Adler:2008gk}. 

We therefore obtain the following equations:
\begin{eqnarray}
\label{eom1}
\left( \omega-i \partial_z -\frac{m_0^2}{2\omega} \right)\phi + 
\frac{g_0 B_T}{2 \omega} \partial_z A_\perp &=& 0 \\ 
\label{eom2}
\left( \omega-i \partial_z  + \Delta_\perp \right)
A_\perp - \frac{g_0 B_T }{2\omega}  \partial_z \phi &=& 0 \\
\label{eom3}
\left( \omega-i\partial_z   + \Delta_{||}  \right) A_{||} &=& 0 \ .
\end{eqnarray}
We have introduced the parameters $\Delta_\perp $ and $\Delta_{||}$ to
take into account that photons may travel in a medium with an index of
refraction $n_i \simeq 1+\Delta_i/\omega$ and again we are assuming
$\Delta_i \ll \omega$.

We notice that only magnetic fields transverse to the direction of
propagation of light lead to mixing effects in this case.  This can be
understood since when the external magnetic field is longitudinal we
have azimuthal symmetry along the propagation axis so the
$z$-component of the angular momentum must be conserved. Since the
photon is helicity-one it cannot convert into a scalar particle.

Let us now turn our attention to the resolution of the equations of
motion. The polarization $A_{||}$ is solved trivially (from now on we
drop the temporal dependence $e^{i\omega t}$, common to all fields):
\begin{equation}
\label{solApar}
A_{||}(z)=e^{-i\omega n_{||} z}A_{||}(0) \ .
\end{equation}
Let us write the equations of the $A_\perp-\phi$ system in matricial
form
\begin{equation}
\label{ }
\left[ \omega - i \partial_z + 
\left(\begin{array}{ccc}  
0   & 
a \\
a &
0
 \end{array} \right)
 (i \partial_z) +
 \left(\begin{array}{ccc}  
- \Delta_0   & 
0 \\
0 &
\Delta_\perp
 \end{array} \right)  \right]
 \left(  \begin{array}{c}   \phi  \\ A_\perp
\end{array} \right) = 0
\end{equation}
where, for convenience, we have performed the shift $A_\perp
\rightarrow i A_\perp$ and we have defined
\begin{equation}
\label{ }
a=\frac{g_0 B_T}{2 \omega} \qquad , \qquad \Delta_0 = \frac{m_0^2}{2 \omega}   \ .
\end{equation}
To solve for the $A_\perp-\phi$ mixing, we search for solutions
$A_\perp(z)={\widetilde A_\perp} e^{-i p z}$ and $\phi(z)={\widetilde
\phi}\, e^{-i p z}$ and are led to the following matricial equation in
Fourier space,
\begin{equation}
\label{matricial_eq}
\left(\begin{array}{cc}  
\omega-p -  \Delta_0   & 
a p \\
a  p  &
\omega-p+ \Delta_\perp 
 \end{array} \right) \
 \left(
\begin{array}{c}  {\widetilde \phi}  \\ {\widetilde A}_\perp
\end{array} \right)
=
 \left( \begin{array}{c} 0 \\ 0
\end{array} \right) \ .
\end{equation}
In order to have non-trivial solutions, the determinant of the matrix
above should vanish. This requirement implies an equation for $p$,
with two solutions: $p_1$ and $p_2$. Introducing the values
$p=p_1,p_2$ in Eq.~(\ref{matricial_eq}) we find that the eigenvectors
$({\widetilde \phi}, {\widetilde A}_\perp)$ corresponding to these
solutions satisfy
\begin{equation}
\label{Thetas}
\frac{{\widetilde A}_\perp^{(1)}}{{\widetilde \phi}^{(1)}}= \frac{-a p_1}{\omega-p_1+\Delta_\perp}\equiv -\tan\Theta_1\,\, ; \,\,
\frac{{\widetilde \phi}^{(2)}}{{\widetilde A}_\perp^{(2)}}= \frac{-a p_2}{\omega-p_2-\Delta_0}\equiv \tan\Theta_2 \ \ .
\end{equation} 
The definition of the angles $\Theta_1$ and $\Theta_2$ is such that
they measure the angular distance (in this flavor space) of the
solution 1 to the pure $\phi$ state $(1,0)$ and of the solution 2 to
the pure photon $(0,1)$ state.

At this level we already find a formal difference with the
pseudoscalar case studied in Ref.~\cite{Raffelt:1987im}.  The
off-diagonal matrix elements of Eq.~(\ref{matricial_eq}) involve the
wavenumber $p$ while in the pseudoscalar case involve the frequency
$\omega$; therefore in general $\Theta_1\neq \Theta_2$, unlike in the
pseudoscalar case~\footnote{Let us stress that this conclusion still
holds even if we do not linearize the equations of motion.}. However,
one can show that
\be
\frac{\tan\Theta_1}{\tan\Theta_2}=
\frac{\omega-\Delta_0}{\omega+\Delta_\perp} \ , 
\ee
and therefore, in the relativistic approximation considered
here~\footnote{When we consider the non relativistic limit $\omega\to
m_0$ (not linearizing the e.o.m. (\ref{eom})) then $\tan\Theta_1\to 0$
while $\tan\Theta_2$ remains finite and the difference between the
scalar and pseudoscalar case is maximal.  This limit has been
carefully discussed for pseudoscalars in Ref.~\cite{Adler:2008gk}.},
we can safely take both angles to be equal.

We will content ourselves showing here the explicit formulas in the
limits
\begin{equation}
\label{limit1}
\Delta_0, \  |\Delta_\perp | \ll \omega \hspace{3.5cm} (\rm relativistic)     
\end{equation}
and
\begin{equation}
\label{weak_mixing}
\Theta_1\simeq \Theta_2 \ll  1
   \hspace{4cm}          (\rm small\ mixing)     
\end{equation}
In this limit the momenta for the two solutions read 
\begin{eqnarray}
p_1 &=& \omega - \Delta_0  - \frac{g_0^2 B_T^2 }{4(\Delta_0 +  \Delta_\perp)}\\
p_2 &=& \omega + \Delta_\perp+\frac{g_0^2 B_T^2}{4(\Delta_0+\Delta_\perp)}\, , 
\label{p}
\end{eqnarray}
where we do not display higher-order terms that are not relevant in
the approximation we work. Introducing these values in
Eq.~(\ref{Thetas}) shows that the solutions $1$ and $2$ are,
respectively, $\phi$-like and photon-like. Let us also define
\begin{equation}
\Theta\equiv  \frac{1}{2}\frac{g_0 B_T }{\Delta_0 +  \Delta_\perp} \simeq \Theta_1\simeq \Theta_2 \ \ .
\label{relation_Fourier} 
\end{equation}
The solution to the linearized equations of motion
Eqs.~(\ref{eom1})-(\ref{eom2}) would be given by a linear combination
of the two found solutions
\begin{eqnarray}
A_\perp(z) & = &  C_1 \,  {\widetilde A}_\perp^{(1)}  \,  e^{-i p_1 z}   \,  
+  \,   C_2  \,   {\widetilde A}_\perp^{(2)}  \,   e^{-i p_2 z} \nonumber \\
\phi(z) & = &  C_1   \,  {\widetilde \phi}^{(1)}  \,   e^{-i p_1 z}  \,   +  \,  
C_2  \,    {\widetilde \phi}^{(2)}  \,   e^{-i p_2 z} \ , \label{general}
\end{eqnarray} 
with the integration constants $C_1$ and $C_2$ determined by the
initial conditions.
  
Now we shall consider the conditions relevant for laboratory
experiments where we start with some finite amplitude $A(0)$ and with
$\phi(0)=0$ and we are interested in the fields after propagating
some distance $z$. Solving for these initial conditions, we get
\begin{eqnarray}
\label{sol-Aperp}
A_\perp(z) &=& A_\perp(0) \,
[ (1 - \Theta^2) \,  e^{-i p_2z} \,  + \,  \Theta^2 \,  e^{-i p_1z} ] \\
\label{sol-phi} 
\phi(z) &=& A_\perp(0)\, \Theta \, [ e^{-ip_2 z}-e^{-ip_1 z} ] \, ,
\end{eqnarray}
where $p_1$ and $p_2$ are given by Eq.~(\ref{p}).

These expressions for the fields can be further simplified when taking
the limit
\begin{equation}
\label{ }
g_0^2 B_T^2  z \ll |\Delta_0  +  \Delta_\perp |   \ .
\end{equation}
We get, at the first non-trivial order in $\Theta$,
\begin{eqnarray}
\label{solAperp}
A_\perp(z) &=&A_\perp(0)e^{-i\omega n_\perp z}\left[
1-\Theta^2 \left(2\sin^2{\frac{b z}{2}}
+ i(b z-\sin{b z}) \right)\right]  \\
\label{solphi} 
\phi(z) &=& A_\perp(0)\, e^{-i\omega n_\perp z} \, \Theta \, [ 1 -e^{ i  b  z} ]
\end{eqnarray}
where 
\begin{equation}
\label{ }
b = \Delta_0 + \Delta_\perp \ . 
\end{equation}

From Eqs.~(\ref{solApar})-(\ref{solAperp}) we see that the different
photon components behave in a different way along their trajectory in
the magnetic field. We can parameterize this difference with a factor
that accounts for the change of the ratio of the amplitudes and a
phase difference as follows:
\begin{equation}
\frac{A_\perp(z)}{A_{||}(z)}=
\frac{A_\perp(0)}{A_{||}(0)}\left(1-\eta(z)\right)e^{-i\varphi(z)} \ ,
\label{defetaphi}
\end{equation}
where $\eta(z)$ and $\varphi(z)$ are real and defined to be
positive. We shall set $n_\| =n_\perp=1$ to focus on the effects of
scalar-photon mixing; in this case $b=\Delta_0$.

The effects of the mixing are a decrease of the amplitude of the wave
$A_\perp$ relative to $A_{||}$ given by
\begin{equation}
1-\eta(z)=1- 2 \Theta^2 \sin^2{\frac{\Delta_0 \, z}{2}} 
\end{equation}
and a phase delay given by
\begin{equation}      \label{phase}
\varphi(z)=\Theta^2(\Delta_0 \,  z -\sin{\Delta_0 \,  z}) \ .
\end{equation}
To this extent, the vacuum filled with a magnetic field acts as a
dichroic and birefringent medium. The consequences for a linearly
polarized wave are: 1) a rotation of its polarization plane and 2) a
small induced ellipticity (as long as the direction of polarization is
not one of the preferred axes, $\perp$ or $||$).

We define the angle of polarization with respect to the external
magnetic field as $\theta=\arctan\frac{|A_\perp|}{|A_{||}|}$. We can
take it in the range $0\leq \theta\leq \pi/2$ since the sign of
the magnetic field does not enter in any observable. Then a small
change of $|A_\perp|$ and/or $|A_{||}|$ will produce a small rotation
$\delta\theta$ given by
\begin{eqnarray}
\label{amplitude}
\delta\theta(z) =\frac{\sin{2\theta}}{2}\left(\frac{\delta
|A_{\perp}|}{|A_\perp|}-\frac{\delta
|A_{||}|}{|A_{||}|}\right)=-\frac{\sin{2\theta}}{2}\eta(z)\ ,
\end{eqnarray}
i.e. the polarization of the laser (its electric field) approaches the
magnetic field plane.  We can understand this as an effect of the
depletion of photons perpendicularly polarized $A_\perp$ with respect
to those polarized along the magnetic field direction $A_{||}$.
Pseudoscalar particles, such as axions, couple instead to $A_{||}$ and
thus $\eta$ would be negative and their production would produce
\emph{positive} rotations of the laser polarization plane, i.e. the
polarization of the laser would tend to be perpendicular to the
magnetic field plane.

The ellipticity $\psi$ of a light wave is defined as
\begin{equation}
\psi=\frac{A_s}{A_l}\ ,
\end{equation}
where $A_s$ ($A_l$) is the shorter (longer) axis of the ellipse that
the vector of the electric field draws in the plane perpendicular to
the propagation. As for the sign, we adopt the same conventions of
Ref.~\cite{Born:1980}, positive ellipticity meaning that the electric
field follows the polarization ellipse in a clockwise sense as seen by
an observer who sees the light propagating towards him.  If the light
is initially linearly polarized then a small phase delay $\varphi$
between $A_\perp$ and $A_{||}$ will lead to a shift of the ellipticity
\begin{equation} \label{defellipticity}
\delta\psi(z)=\delta\left(\frac{A_s}{A_l}\right)=
- \frac{\sin{2\theta}}{2}\varphi(z) \ .
\end{equation}
The ellipticity is anti-clockwise because $A_\perp$ is delayed in time
with respect to $A_{||}$ due to the fact that a small part of the
$A_\perp$ wave travels now with a massive $\phi$-like dispersion
relation.  When this delay takes place $A_{||}$ reaches its maximum a
bit before $A_{\perp}$ when this is still growing. Thus in the maximum
of the amplitude the electric field moves away from the magnetic field
vector.  Notice that if the initial polarization is at an angle of
$\pi/4$ both the rotation $\delta\theta$ and the ellipticity
$\delta\psi$ are maximized.

Since the transition is optimized if the scalar and photon waves are
coherent, we think it is interesting to calculate the rotation and
ellipticity in that limit, which corresponds to $\Delta_0 z \ll
1$. Indeed $ \Delta_0 $ is the momentum difference in vacuum between a
photon and a $\phi$ both with energy $\omega$, which generates, as the
two quanta propagate along the $z$-direction, a relative phase of
$\Delta_0 z$.  Requiring this phase to be negligible is precisely the
coherence condition. In this limit the rotation and ellipticity reduce
to:
\bea
\delta\theta&=& -\frac{1}{16}\, g_0^2 B_T^2 z^2 \, \sin{2\theta} \hspace{2.5cm}
 \rm (coherent)\\
\delta\psi&=& - \frac{1}{96} \, \frac{1}{\omega} \, g_0^2 B_T^2  m_0^2 z^3  \,
\sin{2\theta} \hspace{1.5cm} \rm (coherent) \ .
\eea
As expected, both the rotation and ellipticity have opposite sign with
respect to the ones generated by mixing with pseudoscalars. Except for
this peculiarity, the phenomenology of this mixing is essentially
equal to the axion case.

So far we have considered the case where an initial photon wave
propagates along a magnetic field and is partially converted into a
$\phi$ wave. If we consider the opposite case, i.e. an initially
$\phi$ wave with no photon-component, we expect exactly the same
solutions, Eqs.~(\ref{sol-Aperp})-(\ref{sol-phi}), but interchanging
$A_\perp\leftrightarrow\phi$.  This results can be used to compute the
output of a ``light-shining-through-walls" (LSW) experiment. In such
an experiment a laser beam is shone through a magnetic field onto a
thick wall where photons are stopped but $\phi$'s potentially produced
during the crossing of the magnetic field can traverse.  By the
inverse process, $\phi's$ can be reconverted into photons in another
magnetic field behind the wall and detected in a low background
environment.  Plugging Eq.~(\ref{solphi}) and its equivalent in the
inverse $\phi\to\gamma$ process we find a transmitted wave after the
wall
\be
\label{LSW}
A_\perp(z_c;z_r)= A_\perp(0) e^{-i \omega (z_c+z_r)}
\Theta^2[1-e^{i b z_c}][1-e^{i b z_r}] \, ,
\ee
where $z_c,z_r$ are the lengths of the magnetic fields for the
conversion and reconversion. This leads to a photon regeneration
probability
\be
P(\gamma_\perp\to\phi\to\gamma_\perp)=|A_\perp|^2= 16\, \Theta^4  \sin^2\frac{bz_c}{2} \sin^2\frac{bz_r}{2}  \  ,
\ee
i.e. the same expression as for pseudoscalars when the initial photons
are polarized along the magnetic field direction. Under coherent
conditions this is independent of the $\phi$ mass:
\be
P(\gamma_\perp\to\phi\to\gamma_\perp)=\frac{1}{16}\left(g_0 B_T z_c\right)^2\left(g_0 B_T z_r\right)^2
\hspace{1.5cm} {\rm (coherent)} \, .
\ee


\section{Mixing of Photons with Spin-Two Particles}
\label{tensor}

The Lagrangian describing a free massive spin-two particle $\chi$
was found by Fierz and Pauli in 1939~\cite{Fierz:1939ix}:
\bea
\label{FP}
{\cal L}_{FP}&=&
\frac{1}{4}(\partial_\rho \chi_{\mu\nu})(\partial^\rho\chi^{\mu\nu})-
\frac{1}{2}(\partial_\mu \chi^{\mu\nu})(\partial^\rho \chi_{\rho\nu})+
\frac{1}{2} (\partial_\rho \chi^{\rho\nu})(\partial_\nu \chi^\mu_{\ \mu})\nonumber\\
&& - \frac{1}{4}(\partial_\nu \chi^\mu_{\ \mu}) (\partial^\nu \chi^\mu_{\ \mu})
-\frac{m^2}{4}\chi_{\mu\nu}\chi^{\mu\nu}+\frac{m^2}{4}(\chi^\mu_{\ \mu})^2
\ .
\eea
The equations of motion for $\chi_{\mu\nu}$ can be combined to give
\be
\label{constraints-chi}
\partial_\mu \chi^{\mu\nu}=0 \qquad
\textrm{and} \qquad
\chi^\mu_{\ \mu} =0 \ ,
\ee
which are the usual constraints which apply to the rank-two tensor
describing a spin-two particle, together with the requirement that
$\chi_{\mu\nu}$ is symmetric. We observe that these constraints are
dynamical conditions deriving from the Lagrangian itself.  From the
ten independent entries of a symmetric tensor, conditions
(\ref{constraints-chi}) eliminate five, leaving the five degrees of
freedom that represent the spin-two particle. 

In our study of the mixing we therefore decompose the field $\chi$ as
\be
\label{spin2-dec}
\chi^{\mu\nu}(x) = \sum_i\, \chi_i(x)\,
\epsilon_i^{\mu\nu}(p) \ ,
\ee
where $\chi_i(x)$ are plane-waves and $\epsilon_i^{\mu\nu}$ are
polarization tensors built to satisfy Eq.~(\ref{constraints-chi}).
Here $i$ runs therefore over the five spin-two polarizations:
$+2,\times 2,+1,\times 1,0$.  The explicit formulas for
$\epsilon_i^{\mu\nu}$ are shown in Appendix~\ref{pol}.

Next we consider an interaction Lagrangian of the type
\be
\label{lag-inter}
g \chi^{\mu\nu}O_{\mu\nu}\, ,
\ee
where $O^{\mu\nu}$ is a bilinear in electromagnetic fields and $g$ a
coupling constant.

Starting with the case of parity even spin-two particle, we have in
principle two dimension-four operators candidates for $O^{\mu\nu}$:
$F^{\mu\alpha}F_{\alpha}^{\ \nu}$ and
$\eta^{\mu\nu}F^{\alpha\beta}F_{\alpha\beta}$.  Clearly, the second
couples to the trace of $\chi_{\mu\nu}$, which is zero for our
spin-two particle, and therefore cannot lead to any effect.

On the other hand, if we want our particle to be parity-odd, we find
two analogous candidates. As we demonstrate in
Appendix~\ref{identity}, they are proportional to each other:
\be
F^{\mu\alpha}\widetilde F_{\alpha}^{\ \nu} =-\frac{1}{4}
\eta^{\mu\nu}F^{\alpha\beta}\widetilde F_{\alpha\beta}\ .
\ee
and since one explicitly couples to the trace $\chi^\mu_{\ \mu}$ we
conclude that mixing of photons with parity-odd spin-two particles
should happen via higher-dimensional operators, which will necessarily
include higher orders of the coupling constant $g$.

After this discussion, we are led to considering the mixing of a
parity-even spin-two particle with the following lagrangian
\be \label{lagrangian} {\cal L}= {\cal L}_{EM} + {\cal L}_{FP} +
\frac{g}{2\sqrt{2}}  \chi_{\mu\nu}F^\mu_{\ \alpha}F^{\alpha\nu} \ , \ee
where we have defined the coupling constant with an appropriate numerical 
coefficient to make easier the comparison with the scalar case.

The equations of motion in the Lorentz gauge are, once projected onto the 
spin-two polarizations, 
\bea
(\partial^2+m^2)\chi_i&=&
\sqrt{2} g\epsilon_i^{\mu\nu}{F^{ext}}_{\mu\alpha} ( \partial^{\alpha} A_\nu -
\partial_{\nu} A^\alpha ) \\
\partial^2A^\nu&=&
-\sqrt{2}g \sum_i\ (\epsilon_i^{\mu\alpha}{ F^{ext}}_{\alpha}^{  \
\nu}- \epsilon_i^{\nu\alpha} {F^{ext}}_{\alpha}^{ \ \mu})\ \partial_\mu \chi_i  \ .
\eea
We follow the same procedure as in the scalar case in
Section~\ref{scalar} and shall work under the same conventions and
assumptions.  We linearize and redefine $A_i \rightarrow iA_i$ and we
obtain the following equations in Fourier space:
\bea
\label{chi1x}
&&\left(\omega -p  -\Delta \right){\widetilde \chi_{\times 1}} = 0\\
\label{chi1+}
&&\left(\omega -p  -\Delta \right){\widetilde \chi_{+ 1}} = 0\\
\label{parallel}
&&\left(\begin{array}{cc}  
\omega-p -  \Delta   & 
a_2 p \\
a_2  p  &
\omega-p+ \Delta_\| 
 \end{array} \right) \
 \left(
\begin{array}{c}  {\widetilde \chi_{\times 2} }  \\ {\widetilde A}_\|
\end{array} \right)
=
 \left( \begin{array}{c} 0 \\ 0
\end{array} \right) \\ 
\label{perpendicular}
&&\left(\begin{array}{ccc}  
\omega-p -  \Delta   & 
0 &  a_2 p \\
0 & \omega-p -  \Delta   & a_0 p \\
a_2 p & a_0 p & \omega-p+ \Delta_\perp
 \end{array} \right) \
 \left(
\begin{array}{c}  {\widetilde \chi_{+ 2} }  \\ 
{\widetilde \chi_{ 0} } \\ {\widetilde A}_\perp
\end{array} \right)
=
 \left( \begin{array}{c} 0 \\ 0 \\ 0
\end{array} \right) \ ,
\eea
where we have defined
\begin{equation}
\label{ }
\Delta = \frac{m^2}{2 \omega}
\end{equation}
and
\begin{equation}
\label{a2a0}
a_2 = \, \frac{g B_T}{2 \omega} \qquad , \qquad
a_0 = - \, \frac{g B_T}{\sqrt{3}\, \omega}	\, . 
\ee
This system has seven states that could in principle mix, five for
$\chi$ and two for $A$. However, we notice several interesting
features.  First of all, only the component $B_T$ and the states
$\chi_{\times 2}$, $\chi_{+ 2}$ and $\chi_0$ appear in the mixing
equations: from Eq.~(\ref{chi1x})-(\ref{chi1+}) we see indeed that
$\chi_{\times 1}$ and $\chi_{+ 1}$ decouple. The same argument that we
used in the scalar case can be used here to understand that $B_L$
cannot produce transitions to none of the states $\chi_{\times 2}$,
$\chi_{+ 2}$ and $\chi_0$. In principle $B_L$ could produce
transitions to the $\chi_{\times 1}$ and $\chi_{+ 1}$ (and not $B_T$,
in this case).  However, the states $\chi_{\times 1}$ and $\chi_{+ 1}$
have decoupled and do not appear in the mixing equations. This can be
understood using angular momentum conservation, in a way reminiscent
of the Landau-Yang theorem~\cite{Landau:1948,Yang:1950rg}.  Consider a
massive spin-two particle at rest that decays into two photons and
define the $z$-axis as the direction of propagation of the photons.
It is clear that the spin-two particle cannot be in a spin state
$s_z=+1$ nor $s_z=-1$ because the photons in the final state can only
give $s_z=\pm 2$ or $s_z=0$. By boosting in the $z$ direction, the
same result still holds, so that the particle is decoupled from the
photons if it is in a $\chi_{\times 1}$ and $\chi_{+ 1}$ state.

Furthermore, we see is that the system has decoupled in two blocks,
Eqs.~(\ref{parallel})-(\ref{perpendicular}). This can be understood in
terms of CP symmetry (see Appendix~\ref{pol}). We follow
Ref.~\cite{Raffelt:1987im} and define P as a reflection in the plane
that contains $\vec B$ and the beam (plane $x-z$). The magnetic field
has C$=-1$ and it is a pseudovector, so that CP$=+1$.  The photon
field has also C$=-1$ and the vector character implies CP$=+1$ for
$A_{\perp}$ and C$=-1$ for $A_{||}$. Finally, for the $\chi$ particle,
we have that the polarizations $\times$ correspond to CP odd states,
while $+$ and $0$ to even ones.  All that implies that $A_{\perp}$
couples to $+$ and $0$ while $A_{||}$ to $\times$.  We can now fully
understand the convenience of choosing the $\times$ and $+$
polarizations as we have done.

An interesting result we have obtained is that the $A_{\perp}-\chi_0$
mixing, $a_0$, is of the same order of magnitude as $a_2$ and, in
particular, does not vanish when $m\to 0$.  In
Ref.~\cite{Deffayet:2000pr}, the helicity-0 contribution was
neglected, based on the analogy with the massless case (standard
general relativity) considered in Ref.~\cite{Magueijo:1993ni}.

The structure of the mixing matrix in Eq.~(\ref{perpendicular}) allows
us to perform a simple rotation in the polarization states space of
$\chi$ in such a way that only a linear combination of the spin-two
states couples to the photon.  The combination that couples to
$A_\perp$ is
\be \label{rotated-chi-+} \chi_+ \ =\ \frac{a_2}{a_+}\ \chi_{+2}\  +\
                \frac{a_0}{a_+}\ \chi_0
\ee
with
\begin{equation}
\label{ }
a_+ = \sqrt{a_2^2+a_0^2}\, .
\end{equation}
In terms of this linear combination, the problem reduces to a
two-particle mixing. Instead of Eq.~(\ref{perpendicular}), we have
\begin{equation}
\label{perpendicular2}
\left(\begin{array}{cc}  
\omega-p -  \Delta   & 
a_+\,  p  \\
a_+\,   p  &  \omega-p+ \Delta_\perp
 \end{array} \right) \
 \left(
\begin{array}{c}  {\widetilde \chi_{+} }  \\ 
{\widetilde A}_\perp
\end{array} \right)
=
 \left( \begin{array}{c} 0 \\ 0
\end{array} \right) \ .
\end{equation}
The orthogonal combination $\chi'_+= (a_0 \chi_{+2} - a_2 \chi_0)/a_+$
decouples, so we end up with a simple two-by-two mixing problem for
every photon polarization. The solutions can be read directly from the
scalar case with appropriate momenta and mixing angles. As we said, we
work in the same approximations than in the scalar case, that here
read $\Delta, |\Delta_\||, |\Delta_\perp| \ll \omega$ (already
needed when linearizing the equations of motion) and $g^2
B_T^2/(\Delta + \Delta_i)^2 \ll 1$, for both $i=\|, \perp$.

Using Eq.~(\ref{p}) we find the momenta in the spin-two case:
\begin{eqnarray}
p^{(1)}_\times = 
\omega - \Delta  - \frac{a^2_\times }{b_\times}\, \omega^2
\quad &,& \quad
p^{(2)}_\times = 
\omega + \Delta_\|  + \frac{a^2_\times }{b_\times}\, \omega^2
\nonumber \\
p^{(1)}_+  = 
\omega - \Delta  - \frac{a^2_+ }{b_+}\, \omega^2
\quad &,& \quad
p^{(2)}_+  = 
\omega + \Delta_\perp  + \frac{a^2_+ }{b_+}\, \omega^2
\end{eqnarray}
where, to unify notation, we have defined $a_\times=a_2$ and also
\begin{equation}
\label{ }
b_\times = \Delta + \Delta_\| \qquad , \qquad 
b_+ = \Delta + \Delta_\perp \, .
\end{equation}
The mixing angles are
\begin{equation}
\label{ }
\Theta_\times = \frac{a_\times}{b_\times}\omega \qquad , \qquad
\Theta_+ = \frac{a_+}{b_+}\omega \ . 
\end{equation}
The solutions for ($A_{||},\chi_\times$) and ($A_\perp,\chi_+$) are
completely analogous to Eqs.~(\ref{sol-Aperp})-(\ref{sol-phi}).  In
the limit $g^2 B^2_T z \ll |\Delta + \Delta_i| \ll 1$, for both $i=\|,
\perp$ we have
\begin{eqnarray}
A_\| (z) &=&A_\| (0)e^{-i\omega n_\| z}\left[
1-\Theta_\times^2 \left(2\sin^2{\frac{b_\times z}{2}}
+ i(b_\times z-\sin{b_\times z}) \right)\right]  \\
A_\perp(z) &=&A_\perp(0)e^{-i\omega n_\perp z}\left[
1-\Theta_+^2 \left(2\sin^2{\frac{b_+ z}{2}}
+ i(b_+ z-\sin{b_+ z}) \right)\right] \\
\chi_\times(z) &=& A_\|(0)\, e^{-i\omega n_\| z} \, \Theta_\times \, [ 1 -e^{ i  b_\times  z} ]
\\
\chi_+(z) &=& A_\perp(0)\, e^{-i\omega n_\perp z} \, \Theta_+ \, [ 1 -e^{ i  b_+  z} ]
\\
\chi'_+ (z) &=& 0 \ .
\end{eqnarray}

As in the scalar case, we set $n_\|=n_\perp=1$ to focus on the effects
of photon-$\chi$ mixing so that $b_\times=b_+=\Delta$.  Then, the
relative amplitude change and phase delay of $A_\perp$ and $A_{||}$
are now given by
\begin{eqnarray}
1-\eta(z)=
1- 2 \frac{a_0^2\,\omega^2}{\Delta^2}  \sin^2{\frac{\Delta \, z}{2}} 
\\
\varphi(z)= 
\frac{a_0^2\,\omega^2}{\Delta^2} (\Delta \,  z -\sin{\Delta \,  z}) \ .
\end{eqnarray}
We see that the effects of production of the ${2+}$ and $2\times$
polarizations cancel out~\footnote{This statement turns out to be
slightly modified at the one loop
level~\cite{Bastianelli:2004zp,Bastianelli:2007jv}.}, leaving only the
effect caused by $A_{\perp}\to\chi_0$ transitions.  We could have
expected this since the $A_\perp-\chi_{2+}$ and
$A_{||}-\chi_{2\times}$ mixing \emph{in vacuum} is driven by the same
$a_2$ (see Eqs.~(\ref{parallel})-(\ref{perpendicular})) but $A_\perp$
can also convert into $\chi_0$ and thus depletes and delays its phase
faster than $A_{||}$.

Note that in the massless case $\chi_{2+}$ and $\chi_{2\times}$ are
the only physical components and therefore the cancellation of their
effects does not lead to any net effect neither in the rotation nor
in the ellipticity. However, in the $m\neq 0$ case and even with a
vanishingly small value for $m$, the 0-polarization \emph{does not
decouple} and we expect indeed both effects.  The corresponding
formulas in the coherent case are given by
\bea
\delta\theta&=& -\frac{1}{12}\, g^2 B_T^2 z^2 \, \sin{2\theta}   \hspace{2.5cm}
 (\rm coherent)\\
\delta\psi&=& - \frac{1}{72} \, \frac{1}{\omega} \, g^2 B_T^2  m^2 z^3  \,
\sin{2\theta} \hspace{1.5cm} (\rm coherent) \ , 
\eea
i.e. the same we got for a spin-zero particle, except for a factor
$4/3$ that can be traced back to Eq.~(\ref{a2a0}).  This factor can be
reabsorbed in a redefinition of the coupling constant $g$, so that,
from a positive measurement of $\delta\theta$ and $\delta\psi$ we
would not be able to distinguish between the scalar and the massive
spin-two case.

On the other hand the LSW probability differs from the scalar case,
since here it is non-zero for both photon polarizations. Using the
analogous of Eq.~(\ref{LSW}), in the coherent limit, it is given by
\bea
P(\gamma_\perp\to\chi\to\gamma_\perp)= 
& (a_0^2+a_2^2)^2 \omega^4 z_c^2 z_r^2 &
= \frac{49}{9}\frac{1}{16} \left(g B_T \right)^4 z_c^2 z_r^2      \\
P(\gamma_{||}\to\chi\to\gamma_{||})= 
&a_2^4 \omega^4 z_c^2 z_r^2 & 
=  \frac{1}{16}\left(g B_T\right)^4 z_c^2 z_r^2 \, .
\eea
Note that perpendicular photons have a factor $(1+4/3)^2=49/9\sim 5$
more chances to traverse the wall. Physically this comes from the fact
that the passage through the wall as a $\chi_0$ and as $\chi_{2+}$
adds up coherently at the amplitude level.  


\section{Discussion and Conclusions}
\label{discussion}

To summarize, in this paper we have considered the mixing of photons
with massive scalars $\phi$ and with massive spin-two particles $\chi$
that arises in the presence of a magnetic field. Starting with the
Lagrangian that contains the coupling to two photons, we have
calculated the mixing matrices and the effects on light propagating in
a magnetic field.

The mixing equations of the scalar-photon system present some
differences with the well-studied axion-photon system.  Technically,
this is due to the fact than in the scalar case the interaction term,
i.e. the r.h.s. of Eqs.~(\ref{eom})-(\ref{eom-second}), contains
spatial derivatives of the fields, while in the pseudoscalar case the
derivatives are with respect to time.  This means that we can pass
from the propagation eigenstates in vacuum ($\phi$ and $A_i$) to the
ones in the magnetic field by performing a transformation which is not
a simple rotation. While this introduces a formal difference between
the two cases, in practice this difference disappears when we take the
relativistic limit.  These same issues appear in the spin-two case we
have developed in Section~\ref{tensor}.  Actually, an analogous
situation occurs in the axion-photon mixing in an \emph{electric}
field. In this case the interaction lagrangian is $\propto a\ B \cdot
E$.  While in a constant external magnetic field the axion-photon
mixing results to be $\propto \partial_0 A\cdot B_{\rm ext} = i \omega
A_{||} B_{\rm ext}$, in an external electric field is $\propto (\nabla
\times A)\cdot E_{\rm ext}=-i p A_\perp E_{\rm ext}$, in analogy with
the case discussed here.

When constructing the interaction Lagrangian, we have seen that the
interaction of a $2^+$ particle is a dimension-five operator, as
expected. However, the interaction operators in the $2^-$ case are at
least dimension-seven, so that they will be much more suppressed.  For
this reason we have only calculated the parity-even case.

Referring now to the mixing of a $2^+$ particle with photons, we have
seen that the helicity-one states $\chi_1$ decouple.  This can be
understood with arguments of rotational symmetry, as explained in
Section~\ref{tensor}. Apart from the decoupling just mentioned, there
is further decoupling since $A_{||}$ couples only to the $\times 2$
mode while $A_\perp$ couples only to the $+ 2$ and 0 modes. As also
explained in Section~\ref{tensor}, it is a consequence of CP symmetry.
We have seen that the $A_\perp-\chi_2$ and the $A_{||}-\chi_2$ mixing
have identical value, but $\chi_0$ makes a difference: it mixes with
$A_\perp$ only.  As a consequence, the contributions to the rotation
and ellipticity observables of $\gamma-\chi$ mixing contain only the
$A_\perp-\chi_0$ mixing.  Moreover, compared to the $A-\chi_2$ mixing
amplitude, the $A-\chi_0$ mixing is {\em not} suppressed, indeed it is
larger of a factor $\sqrt{4/3}$.

In the eventual discovery of effects of rotation and ellipticity in
light propagating in a magnetic field, and in the eventual case that
the signs would correspond to a parity-even particle, the tensor case
should be considered as a possible explanation, together of course
with the scalar case.  In such a case, a
``light-shining-through-walls" experiment would be useful to
discriminate between these two candidates since photons with
polarization parallel to the magnetic field would lead to a signal
\emph{only in the spin-two case}.  This is because, in this case, both
polarizations would lead to positive signals.  Moreover, we have found
that the probability for photons polarized perpendicularly to the
magnetic field is a factor $(1+4/3)^2$ larger than for photons with
parallel polarization, due to the additional $\chi_0$ intermediate
state.

While in the massless spin-two case there are no observable effects of
rotation and ellipticity, we have found that there are effects when
taking the $m \rightarrow 0$ limit of the massive theory because the
$A-\chi_0$ mixing does not vanish.  Our particle could be a massive
graviton but it could be as well a tensorial particle with no relation
whatsoever with gravitation.  This result might be of theoretical
interest in the light of the controversy outlined in
Section~\ref{introduction}.

%


\section*{Acknowledgements} We would like to thank D.~Blas, E.~{\'A}lvarez and P.~Silva for
discussions. We acknowledge support by the CICYT Research Projects
FPA2003-04597 and FPA2005-05904, the \textit{Departament
d'Universitats, Recerca i Societat de la Informaci{\'o}} (DURSI),
Project 2005SGR00916 and the European Commision under the RTN program
MRTN-CT-2004-503363. C.B. is grateful to IFAE (Universitat
Aut{\`o}noma de Barcelona) for hospitality while part of the work was
performed.


\appendix
\section*{Appendixes}


\section{Polarization of a spin-two particle}
\label{pol}

It has been known since a long time~\cite{auvil} how to construct the
polarization basis of particles with arbitrary spin starting from spin
one-half and one. In particular spin-two can be built from spin-one.
For a vector particle propagating along $z$ with four-momentum $P^\mu=
(\omega;0,0,p)$, the helicities $+1$, $-1$ and $0$ can be chosen to be
\begin{eqnarray}
\epsilon^\mu(h = \pm 1) &=& (0, \mp 1, -i, 0) \nonumber \\
\epsilon^\mu(h = 0) &=& (p/m,  0, 0, \omega/m)\ .
\end{eqnarray}
where here $m$ has to be implicitly understood as $m=\sqrt{P_\mu
P^\mu}=\sqrt{\omega^2-p^2}$.

By using the appropriate Clebsch-Gordan coefficients, one can obtain
the helicities $h=0,\pm1,\pm2$. For example,  $h=+2$ is given by
\begin{eqnarray}
\epsilon^{\mu\nu}(h=+2) &=&\epsilon^{\mu}(h=+1)\epsilon^{\nu}(h=+1)=
\frac{1}{2}\ \left(
\begin{array}{cccc}
0 & 0 & 0 & 0 \\
0 & 1 & i & 0 \\
0 & i & -1 & 0 \\
0 & 0 & 0 & 0
\end{array}
\right) \end{eqnarray}
and so on. For our calculations it is more convenient to work with
CP-eigenstates. Since CP transforms $+h$ into $-h$, we will use the
following combinations
\begin{eqnarray}
\epsilon^{\mu\nu}_{+2}\ &=&
  \frac{-1}{\sqrt{2}}(\epsilon^{\mu\nu}(h=+2)+\epsilon^{\mu\nu}(h=-2))=
           \frac{1}{\sqrt{2}}\left(\begin{array}{cccc}
                         0 & 0 & 0 & 0 \\
                         0 & -1 & 0 & 0 \\
                         0 & 0 & 1 & 0 \\
                         0 & 0 & 0 & 0 \\
                         \end{array}\right)
\end{eqnarray}
\begin{eqnarray}
\epsilon^{\mu\nu}_{\times 2}\ &=&
\frac{1}{\sqrt{2}i}(\epsilon^{\mu\nu}(h=+2)-\epsilon^{\mu\nu}(h=-2))=
           \frac{1}{\sqrt{2}}\left(\begin{array}{cccc}
                         0 & 0 & 0 & 0 \\
                         0 & 0 & 1 & 0 \\
                         0 & 1 & 0 & 0 \\
                         0 & 0 & 0 & 0 \\
                         \end{array}\right)
\end{eqnarray}
\begin{eqnarray}
\epsilon^{\mu\nu}_{+1}\ &=&
  \frac{1}{\sqrt{2}i}(\epsilon^{\mu\nu}(h=+1)+\epsilon^{\mu\nu}(h=-1))=
           -\frac{1}{m\sqrt{2}}\left(\begin{array}{cccc}
                         0 & 0 & p & 0 \\
                         0 & 0 & 0 & 0 \\
                         p & 0 & 0 & \omega \\
                         0 & 0 & \omega & 0 \\
                         \end{array}\right)
\end{eqnarray}
\begin{eqnarray}
\epsilon^{\mu\nu}_{\times 1}\ &=&
  \frac{1}{\sqrt{2}}(\epsilon^{\mu\nu}(h=+1)-\epsilon^{\mu\nu}(h=-1))=
            \frac{1}{m\sqrt{2}}\left(\begin{array}{cccc}
                         0 & p & 0 & 0 \\
                         p & 0 & 0 & \omega \\
                         0 & 0 & 0 & 0 \\
                         0 & \omega& 0 & 0 \\
                        \end{array}\right)
\end{eqnarray}
\begin{eqnarray}
\epsilon^{\mu\nu}_0\  = \epsilon^{\mu\nu}(h=0)= \label{epsilon0}
\frac{1}{m^2}\sqrt{\frac{2}{3}}
 \left(\begin{array}{cccc}
                         2p^2 & 0 & 0 & 2p\omega \\
                         0 & -m^2 & 0 & 0 \\
                         0 & 0 & -m^2 & 0 \\
                         2p\omega & 0 & 0 & 2\omega^2 \\
                         \end{array}\right)\ .
\end{eqnarray}
Note that, as expected, for all $i=0,+ 1,\times 1,+ 2, \times 2$ the
requirements of Eq.~(\ref{constraints-chi}) are satisfied,
\begin{equation}
P_\mu\epsilon_i^{\mu\nu} =0 \hspace{1.7cm} {\epsilon_i}^\mu_{\ \mu}
=0 \hspace{1.7cm} \epsilon_i^{\mu\nu}=\epsilon_i^{\nu\mu}
\end{equation}
and also
\begin{equation}
\epsilon_{\mu\nu\, i}\, \epsilon^{\mu\nu}_{\ j}=\delta_{ij} \quad \forall i,j \ .
\end{equation}

\section{ A useful identity}
\label{identity}

Given $A_{\mu\nu}$ and $B_{\mu\nu}$ antisymmetric we first evaluate
\begin{equation}
\widetilde A_{\mu\nu} \widetilde B^{\mu\rho} = \frac{1}{4}\
\epsilon_{\mu\nu\alpha\beta}\  \epsilon^{\mu\rho\sigma\tau}\
A^{\alpha\beta}\ B_{\sigma\tau} \ .
\end{equation}
Then we use
\begin{equation}
 \epsilon_{\mu\nu\alpha\beta}\  \epsilon^{\mu\rho\sigma\tau} =
 \left| \begin{array}{ccc}
 \delta^\rho_\nu & \delta^\rho_\alpha & \delta^\rho_\beta \cr  \\
 \delta^\sigma_\nu & \delta^\sigma_\alpha & \delta^\sigma_\beta \cr \\
 \delta^\tau_\nu & \delta^\tau_\alpha & \delta^\tau_\beta 
 \end{array}
 \right|
\end{equation}
to obtain
\begin{equation}
\widetilde A_{\mu\nu} \widetilde B^{\mu\rho} = \frac{1}{2}\
\delta^\rho_\nu\ A_{\alpha\beta} B^{\alpha\beta} \ - \
A^{\alpha\rho} B_{\alpha\nu}\ .
\end{equation}
We apply now the identity to $\widetilde A=F$ and $\widetilde
B=\widetilde F$:
\begin{equation}
F_{\mu\nu} \widetilde F^{\mu\rho} = \frac{1}{2}\ \delta^\rho_\nu\
\widetilde F_{\alpha\beta} F^{\alpha\beta} \ - \ \widetilde
F^{\alpha\rho} F_{\alpha\nu}\ ;
\end{equation}
since the last term is identical to the l.h.s., we finally obtain:
\begin{equation}
F_{\mu\nu} \widetilde F^{\mu\rho} = \frac{1}{4}\ \delta^\rho_\nu\
\widetilde F_{\alpha\beta} F^{\alpha\beta}\ .
\end{equation}


\providecommand{\href}[2]{#2}\begingroup\raggedright\endgroup


\end{document}